\input{aipcheck}

\documentclass[final,numberedheadings]{aipproc}

\layoutstyle{8x11single}

\begin{document}

\title{Acquisition of Information is Achieved by the Measurement Process in Classical and Quantum Physics
\footnote{Prepared for : ''{\textsc{Quantum Theory}: \emph{Reconsideration of Foundations}$-4$}'' (QTFR$-4$), V\"{a}xj\"{o}, Sweden, $6-11$ June $2007$. To be published by the American Institute of Physics in the AIP Conference Proceedings series. Talk presented by \textsc{Paolo Rocchi}.}
}


\author{Paolo Rocchi}{
  address={IBM, Via Shangai 53, Rome, Italy}}

\author{Orlando Panella}{
  address={INFN, Sezione di Perugia, Via A. Pascoli, I-06123, Perugia, Italy}
}


\date{October 24, 2007}

\begin{abstract}
No consensus seems to exist as to what constitutes a measurement which is still considered somewhat mysterious in many respects in quantum mechanics. At successive stages \emph{mathematical theory of measure, metrology} and \emph{measurement theory} tried to systematize this field but significant questions remain open about the nature of measurement, about the characterization of the observer, about the reliability of measurement processes etc. The present paper attempts to talk about these questions through the information science. We start from the idea, rather common and intuitive, that the measurement process basically acquires information. Next we expand this idea through four formal definitions and infer some corollaries regarding the measurement process from those definitions. Relativity emerges as the basic property of measurement from the present logical framework and this rather surprising result collides with the feeling of physicists who take measurement as a myth. In the closing this paper shows how the measurement relativity wholly consists with some effects calculated in QM and in Einstein's theory.
\end{abstract}

\maketitle


\section{Introduction}
Normally measurement is taken as an operational sector whose major scope is to support scientists in different territories, and also professionals in various areas: sellers, workers, engineers etc.  Some attempts have been made to provide a comprehensive framework for measurement. The term metrology has been coined to define a novel science which should embrace both experiment and theoretical determinations at any level of uncertainty in any field of science and technology. Despite this pompous manifest, in substance metrology defines, realizes and disseminates units of measurement~\cite{Harlow2003}. The abstract theory of measure involved eminent authors (Borel, Lebesgue, Lyapunov, Kolmogorov, Carathéodory etc.) but does not justify the inner nature of measure which remains unclear. Modern books on measurement theory show the profile of a pragmatic discipline and include statistical methods, construction of instruments and probes, operations for measurements, laws and legal procedures, and other empirical topics~\cite{Gertsbakh2003}. In conclusion, the clear understanding of what a measure consists appears not close.
Several authors agree that information strongly affects the nature of a measure and we assume that the measurement process basically consists of information acquisition. In this way we shall examine the most significant features of measure through the lens provided by theoretical works on information, and try to answer some questions risen by philosophers and by theorists in classical and quantum physics, such as: May systematic errors of measure be erased? Why and how does a measurement process influence the outcomes? What is the observer characterization? Does the observer lie outside or inside a physical test? Is the intellectual awareness a part of the quantum measurement process?

\section{Beyond the jungle}
Various scientific theories about information have been put forward and stand
often without the least explicit relationship to each other. We see so many
differing definitions that also the classification of theories appears a
tricky question~\cite{Floridi2004}. The ensuing incomplete list gives an idea about the
variety of schools:
\begin{itemize}
\item[{-}]  Semantic theory of information (Bar-Hillel and Carnap)
\item[{-}]  Algorithmic theory of information (Solomonoff, Kolmogorov, Chaitin),
\item[{-}]  Qualitative theory of information (Mazur),
\item[{-}]  Economical theory of information (Marschak),
\item[{-}]  Social theory of information (Goguen),
\item[{-}]  Autopoietic theory of information (Maturana and Varela),
\item[{-}]  Statistical theory of information (Shannon).
\end{itemize}

No adequate concept as well as understanding of information has been produced by theoretical researchers over the last decade of the 20th century~\cite{Israel:1990}. The cultural scenario is so much confusing that theoretical physicists, who aim at exploiting the notions of the information and communication technology (ICT), normally cannot go ahead. The lack of a solid reference appears like a huge rock amid the route that obstructs the progress unless one finds a way to circumvent this barrier.

The present paper suggests getting around this obstacle by means of multidisciplinary conceptualization.

\section{MULTIDISCIPLINARY APPROACH}

Several disciplines, which revolve around the notion of information, deal with a broad variety of problems, nonetheless we can identify two principal views shared by authors. A first group (e.g. computer hardware, networking, neurophysiology) experiences the physical nature of information, namely they investigate the material form of information and its properties. A second group of disciplines (e.g. computer software, semiotics, logic, linguistics, psychology) communalize the semantic view to information namely they focus on the contents of texts, of messages, etc.

Quantum theorists are increasingly aware of the influence of cognitive sciences on the measurement problem and we take the perspective which integrates Shannon's work - on which technicians and engineers prevalently ground their theoretical studies - with the humanities. We specify this multidisciplinary way to information and to measure by means of the following principles which sum up the perspectives of the above mentioned two groups:

\vspace{0.15cm}

\noindent\textbf{Principle 1}:   \emph{A piece of information is a physical distinguishable item}.\\
\noindent \textbf{Principle 2}:   \emph{Information finds its definitive significance with respect to the human knowledge}.\vspace{0.15cm}

We explode the foregoing principles in the following way.
A piece of information is present from background events and the receptor R is capable of detecting this piece if it is neat. The generic stuff $E$, whether an artifact (i.e. writings, pixels) or natural element (i.e. a quantum particle), is capable of informing on condition it is distinguishable. The entity $E$ is impossible to detect save as $E$ contrasts with an adjacent entity $E^*$. Discriminability is the special feature of all the pieces of information which we define in the following way.\\

\noindent\textbf{Definition 1}: \emph{The entity E is said to be an item of information if E is distinct from a given comparison-entity E* with respect to the reference R}
\begin{equation}
E_R = NOT\, E^*\end{equation}
\emph{Where  $E$ and $E^*$ are entities of the algebraic space ${\cal E}$.}

The vast majority of authors in semiotics and in linguistics accept that the item of information E replaces the object, the event or the idea $NE$; to wit $E$ stands for $NE$. The former is frequently named \emph{signifier} and the latter \emph{signified}. We write the \emph{signification} between the signifier $E$ and the signified $NE$, adopting the Lacan symbolism~\cite{Lacan:1977}.\\

\noindent\textbf{Definition 2}:  \emph{The entity E  refers to the entity NE}
\begin{equation}
\frac{E}{NE}
\end{equation}

From Definitions 1 and 2 we infer that the system for the acquisition of information includes two blocks: the former detects E and operates without any intelligence, the latter assigns the significance to the readout.\\

\noindent\textbf{Definition 3}:  \emph{The system to acquire information is equipped with the detection process R and the semantic process S}.
\begin{figure}[h!]
  \includegraphics[height=.08\textheight]{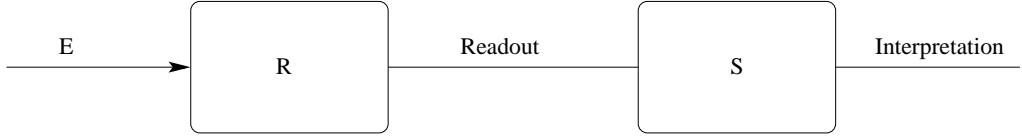}
  \caption{The $R-S$ system}
\end{figure}
Information takes origin from sensation, the mechanical and initial act to acquire a sign, and then from the recognition stage $S$ which determines the meaning of the acquired signal. In practice $R-S$ is an artifact (i.e. measurement tool, detector), a biological system such as the human observer or a combination of both. Experience shows how the stages $R$ and $S$ may be generic, fuzzy and subjective due to human intervention, thus we define the most accurate conditions for $R-S$.\\

\noindent\textbf{Constraint 1}:  \emph{We assume $E$, $E^*$ are points in a metric mono-dimensional space, and from (1) we derive the condition of distinction}
\begin{equation}
E\neq E^*\, .
\end{equation}

\noindent\textbf{Constraint 2}:  \emph{We assume the signified NE is a standard unit, compatible with E}
\begin{equation}
NE = EU  \, .\end{equation}

$E$ and $EU$ are exact and comparable entities due to (3) and (4), thus the semantic relation (2) becomes a ratio which provides a pure number

\begin{equation}
\frac{E}{EU}= Em
\end{equation}
We conclude that $(Em\cdot EU)$ quantifies E to wit it provides the measure of $E$.\\

\noindent\textbf{Definition 4} : \emph{The following product is said the measure of $E$
}
\begin{equation} Em \cdot EU \end{equation}

The precision of the number $Em$ gives the granularity of the measure and $EU$ provides the meaning of the measure, thus (6) consists with Definitions 1 and 2.

The parallel Principles 1 and 2, the coupled Definitions 1 and 2, the operations $R$ and $S$, the pair (6) show that measure has a double nature: physical and semantic. The dual origin of measure, in particular its reliance on the observer who establishes the interpretation of $E$, seemed a kind of scientific taboo~\cite{Bub:2005}. Theorists frequently treat the human intervention in generic terms, because they fear the subjectivism and the generic behavior of the humans instead here the observer $R-S$ finds a rigorous formalization.

The present paper does not develop the mathematics from the foregoing definitions; instead we shall discuss the major properties of measurement that is to say we go deep into the nature of measurement in order to improve our understanding of this topic.

\section{RELATIVISM: THE BASIC PROPERTY OF MEASURE}
Definitions 1 and 2 include the following set of arbitrary elements which experimentalists factually establish at their convenience
\begin{equation}
{E^*, R, NE, EU}\, .\end{equation}

In particular the reference is chosen so that $R$ is close and physically compatible to $E$ otherwise $R$ cannot achieve the detection process. The comparison term $E^*$ is selected so that $E$ can contrasts. Sometimes the interpretation $NE$ is a figment of human imagination and the semantic relationship does not follow any rule. Scientists determine the standard unit $EU$ under the influence of sociological and political issues. In conclusion the elements (7) works out as essential parts for the acquisition of information but the present theory does not regulate them in terms of mathematics. The elements (7) are subjected to empirical criteria and not to the necessity of general reasons, thus we conclude that \textbf{measure is a relativistic quantity}. Relativism emerges as the fundamental property of a measure in the light of the present logical framework hence this seemingly paradoxical achievement is to be expanded.

The arbitrary items (7) bring about four genders of relativism which we call this way:
\begin{itemize}
\item[{-}]  \emph{Couple relativity due to E*},
\item[{-}]  \emph{Reference relativity due to R},
\item[{-}]  \emph{Semantic relativity due to NE},
\item[{-}]  \emph{Unit relativity due to EU}.
\end{itemize}

Current literature amply discusses the doctrines of semantic arbitrariness as semantic relativity affects natural languages~\cite{Culler:1986}. Physicists are aware that the selection of units influences the arrangement of systems of measurement. The criteria followed to fix sets of units which must be used to specify anything are illustrated in the domain of experimental sciences~\cite{Anthony:1986}. The couple relativity regards the comparison term which does not produce heavy effects~\cite{Rocchi:2006}. Finally the detector $R$ causes the most demanding consequences in physics which we discuss next.

\subsection{The Reference Relativity Cannot be Developed within a Unified Theory}

Eq. (1) establishes a functional relationship between $E$ and $E^*$
\begin{equation}
E=f(E^*)
\end{equation}

Thus the couple relativity may be treated within a unified mathematical frame [9], instead the reference relativity cannot be treated because of the generic relationship between $E$ and $R$ in (1)
\begin{equation}
E =  E_R
\end{equation}

A broad variety of phenomena which inflict damages upon measurements brings evidence of this corollary. The eclectic nature of the reference relativity emerges from the following cases randomly selected from classical physics.
\begin{itemize}

\item[{1)}] The sensor $R$ works within the minimum threshold $\min t_R$ and the maximum threshold $\max t_R$. If $E$ does not fall in the available interval ($\min t_R$, $\max t_R$), the measure of $E$ does not exist.
\item[{2)}] The sensor may to some extent be sensitive for other properties than the property being measured. To exemplify, most instruments are influenced by the temperature of their environment.
\item[{3)}] If the output signal slowly changes independent of the measured property, this is defined as \emph{drift}. Long term drift usually indicates a slow degradation of sensor properties over a long period of time.
\item[{4)}] Noise is a random deviation of the signal that varies in time. Noise can be reduced by signal processing, such as filtering, usually at the expense of the dynamic behavior of the sensor.
\item[{5)}] Hysteresis is an error caused by the fact that the sensor not instantly follows the change of the property being measured, and therefore involves the history of the measured property.
\end{itemize}

Errors, as from 1 to 5, make visible the reference relativity in classical physics. The efforts to compensate those errors by means of some kind of calibration strategy reduce the impact of those errors, but do not deny the reference relativity in principle.

In quantum field theoretical principles sustain the study of the measurement limits~\cite{Healey:1998}.

\begin{itemize}
\item[{A)}] As first we quote the \emph{uncertainty principle} by means of the Robertson-Schr\"{o}dinger inequality. Given the Hermitian operators $A$ and $B$ and a system in the state $\psi$, there are probability distributions associated with the measurement of each $A$ and $B$, giving rise to standard deviations $(\Delta A)_\psi$ and $(\Delta B)_\psi$, then

\begin{equation}
(\Delta A)_\psi \, (\Delta B)_\psi \geq \frac{1}{2} \left| \langle [A,B]\rangle_\psi \right|
\end{equation}
Where the operator $[A,B] = AB - BA$  is the commutator of $A$ and $B$,
and $\langle X \rangle_\psi$ denotes the expectation value. Due to this principle an uncertainty relation arises between any two observable quantities that can be defined by non-commuting operators e.g. position and momentum of a particle; angular position and angular momentum of a particle. The uncertain principle regards also the pair time-energy.

\item[{B)}] Quantum particles are actually waves with complex amplitude spread out in space and evolving according to indeterministic laws, except when a measurement occurs. During a measurement, the wave-function instantaneously changes into a state consistent with one of the possible measurement outcomes. The probability of any given outcome is dependent on the amplitude of the wave. In most cases, this means that the wave-function becomes suddenly localized in space, when before it was widely distributed. We express the state of a quantum system
\begin{equation}
|\Psi\rangle = \sum_i |i\rangle \psi_i
\end{equation}
Where $|i\rangle$ form an orthonormal eigenvector basis. The $\psi_i = \langle i |\Psi\rangle $ are the probability amplitude coefficients which we assume are normalized. When an observer measures the observable associated with the eigenbasis then the state of the wave function changes from $|\Psi\rangle$ to just one of the $|i\rangle$'s with probability $|\psi_i|^2$. All the other terms in the expansion of the wave function vanish into nothing due to the observer $R-S$. The act of measurement itself determines the state of the system as it is measured.
\item[{C)}] When a quantum state is entangled (i.e. two particles) it is possible to obtain the value of an observable of the particle $E_2$ with a measurement performed on particle $E_1$. The EPR \emph{paradox} deals with the related violation of locality in ordinary QM in these types of experiments. In this case an observer performs measurement only on a particle but the state of the other is determined in a total indirect way. Let the observer $R_{E_2}$ is located much further from the birthplace of the particle $E_1$ and $E_2$ than the observer $R_{E_1}$. Let the observer $R_{E_1}$  measures the projection of spin into the axis $z$ for the particle $E_1$, and this value appears as $S_z(E_1)$. The particles $E_1$ and $E_2$ were
    in a singlet state at once after the decay of the object Q. Therefore the observer $R_{E_2}$ will find out a value
\begin{equation}
S_z(E_2) = S_z(E_1)
\end{equation}

For the particle $E_2$ with probability equal to the unity. Within the framework of standard quantum mechanics it means that the particle $E_2$ is in a state with definitive value of projection of spin $S_z(E_2)$ on the axis $z$.
\end{itemize}

Einstein forecasts some special effects on measurement which directly derive from the special relativity theory and the general relativity theory~\cite{Brown:2005}.
\begin{itemize}
\item[{I)}] Take an event with space-time coordinates $(t,x,y,z)$ in system $R$ and $(t',x',y',z')$ in $R'$ which is moving at the speed $v$. The Lorentz transformation specifies that these coordinates are related in the following way
    \begin{equation}
    \begin{array}{ll}
    t'= \gamma\left(t-\displaystyle\frac{v}{c^2}x\right) & y'=y\\
    x'=\gamma(x-vt)& z'=z
    \end{array}
    \end{equation}
where $\gamma $  is the Lorentz factor
\begin{equation}\gamma=\frac{1}{\sqrt{1-\frac{v^2}{c^2}}}
\end{equation}
The time lapse between two events is dependent on the relative speeds of the observers' reference frames. In fact we write the Lorentz transformation and its inverse in terms of coordinate differences and obtain
\begin{equation}\Delta t' =\gamma\left(\Delta t -\frac{v}{c^2}\,\Delta x \right)\end{equation}
Take as an example two consecutive ticks of a clock characterized by  $\Delta x=0$, the relation between the times of these ticks as measured in both systems is given by
\begin{equation} \Delta t'=\gamma \Delta  t \end{equation}
This equation shows the time dilatation, namely the time $\Delta t'$ between the two ticks as seen in the \emph{moving} frame $R'$ is larger than the time $\Delta t $ between these ticks as measured in the rest frame of the clock.

\item[{II)}]    Suppose the length of an object $\Delta x$ in the frame $R$, and we calculate the length of the same object in $R'$. Assuming that the end points are measured simultaneously in $R'$, namely  $\Delta t'=0$, we obtain using the Lorentz transform
\begin{equation}
\Delta x' =\frac{\Delta x}{\gamma}\, .
\end{equation}
This shows that the length contraction to wit length $\Delta x'$ of the object measured in the moving frame $R'$ is shorter than the length $\Delta x$ in its own rest frame.
\item[{III)}]   The mass of an object (also known as relativistic mass) increases with the speed according to the following expression
\begin{equation}
M =  \gamma m,
\end{equation}

where $m$ is the invariant mass, and $\gamma$  the Lorentz factor.

\item[{IV)}]    According to Einstein's general relativity the rotation of
 object alters space and time, dragging a nearby object out of position
 compared to the predictions of Newtonian physics. In other words an
 observer who is distant from a rotating massive object and at rest with
 respect to its center of mass will find that the fastest clocks at a
 given distance from the object are not those which are at rest
 (as is the case for a non-rotating massive object).
 Instead, the fastest clocks will be found to have component
 of motion around the rotating object in the direction of the
 rotation. This effect named \emph{Lense-Thirring effect} or \emph{frame
 dragging} occurs where a rotating massive object "drags" space-time
 along with its rotation. For example frame dragging causes the
 orientation of a gyroscope to change over time.
\end{itemize}

Empirical case studies justify effects 1-5, by contrast effects A-C
derive from a theoretical frame not yet unified in QM. Einstein
calculates effects I-IV which directly derive from the special
relativity theory and from general relativity theory. This means
that points 1-5, A-C and I-IV provide support, progressively
stronger, to the reference relativity of measure. Three intellectual
constructions - empirical methods, quantum theory, Einstein's theory
- converge toward the same issue: \textbf{measurement is not an
absolute process} due to the influence of arbitrary entities - in
particular R. This conclusion seems to clash against the universal
confidence on measurement as only measured results can validate a
scientific theory. Physicists tend to refuse the relativism of
measurement - evident in quantum mechanics - because they fear the
scientific method may be taken unworthy. Instead this suspect is
misleading because the measurement relativity derives from the
precise set (7). We are capable of analyzing each arbitrary element
and in consequence are capable of calculating its actions. If an
experimentalist is unable to neutralize a distortion or a limitation
caused by an arbitrary element in the practice, anyway he can
forecast this negative effect with precision. E.g. the wave collapse
restricts an experimental inquiry anyway the result obtained after
the collapse is valid because we are aware of this effect. We can
reasonably conclude that the measurement relativity - especially the
reference relativity - does not entail that the scientific method
deviates the human mind but simply holds that the scientific method
is difficult to apply due to the relativistic effects.

\subsection{Where the Observer Lies}
The problems of inclusion/ exclusion of the observer from the
physical experiment assume special significance because Principle 2
holds that the observer is necessary, moreover we have seen how the
observer $R-S$ causes additional effects. The reply to the question
does not seem easy as it is insufficient to hold that the observer
lies outside or inside the experiment.
\begin{figure}[h!]
\includegraphics[height=0.2\textheight]{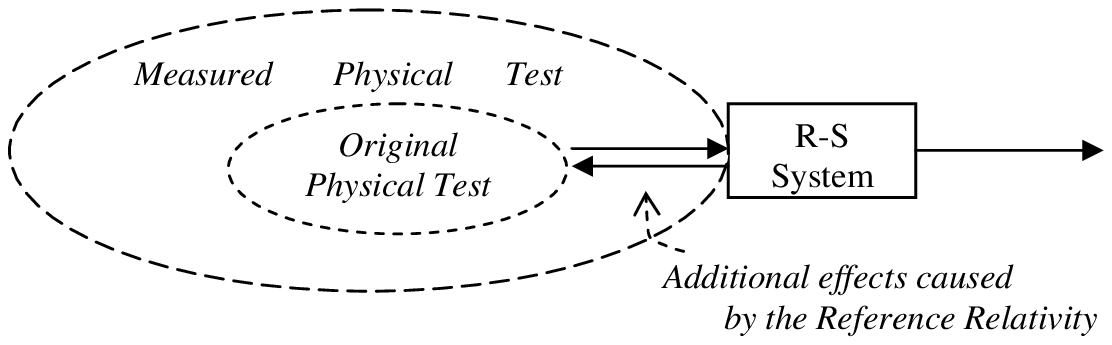}
  \caption{}
\end{figure}

We have exactly determined the operations of $R-S$ thus we are able
to divide the \emph{original test} from \emph{the measured test}
that includes the interactions provoked by R-S. Therefore an
observer lies outside the initial experiment and lies inside the
measured experiment because of his interactions. This statement
reinforces the notion that measurement is not neutral and the
measurement relativity takes essential role.

The present corollary somehow agrees with the thought expressed by
von Neumann, David Bohm, Henry Stapp, Freeman Dyson, and others who
claim "consciousness causes collapse"\cite{Bohm:1980}. In fact the
consciousness may be seen as a part of the block S which makes an
essential component of an observer.

\subsection{Quantum Decoherence: an Aspect of the Measurement  Relativity}

The collapse constitutes the special and partially mysterious interference between a particle and the measurement system. There are two major schools toward the wave function collapse. The former rejects the collapse as a physical process and relates to it only as an illusion. Those who support this approach usually offer another interpretation of quantum mechanics, which avoids the wave function collapse. The latter schools supported by Niels Bohr and his Copenhagen interpretation accept the collapse as one of the elementary properties of nature. The present conceptualization matches with the last authors in particular it agrees with physicists who aim at identifying the fuzzy boundary between the quantum micro-world and the world where the classical intuition is applicable. The interaction between the microscopic level of the particle $E$ and the macroscopic level of the system $R-S$ appears as a special form of reference relativity which may be treated by means of the decoherence theory. Quantum decoherence was proposed in the context of the many-worlds interpretation, but it has also become an important part of some modern updates of the Copenhagen interpretation. Technically decoherence consists in detailing a model for the system-environment interactions and in deriving master equations for the reduced density matrix of the system.\\

\noindent{\emph{Density matrix description of decoherence.}}\\
When studying a point mass the inclusion of the unavoidable scattering processes with the particles of the environment gives the non-unitary part of the time evolution of the density matrix. In the coordinate representation the master equation for the density matrix reads
\begin{equation}
i\hbar  \frac{\partial \rho(x,x')}{\partial t} = \langle x| [H,\rho] |x' \rangle -i\hbar \rho(x,x') F(x-x')
\end{equation}

Where the quantity $F(x-x')$ can be connected to the scattering cross-section of the point mass. The unitary evolution based on the Schr\"{o}dinger equation gives only the first term in the previous equation. Based on the assumptions that: 1) the recoil in the scattering is negligible  (only the state of the scattered particle is changed during the interaction); 2) very many particles are scattered during a short time interval and 3) the incoming particles are distributed isotropically, one finds that $F(x-x')= \Lambda (x-x')^2$  with
  $\Lambda =k^2 n v \sigma_{eff}$ (called the localization rate, measured in cm$^2$/sec) where $k$ is the momentum of the scattering particles, $n$ their density and $\sigma_{eff}$  their effective cross section with the point mass. Solving the master equation yields the suppression of off-diagonal terms of the reduced density matrix  i.e.  decoherence
\begin{equation}
\rho(x,x',t) =\rho(x,x',0)\, \exp[-\Lambda t(x-x')^2 ]
\end{equation}
See these references~\cite{Joos:1984uk,PhysRevA.42.38,PhysRevD.34.470} for a review of explicit cases.\\

\noindent \emph{Wigner function description of decoherence}\\
A commonly used representation of both pure and mixed states, alternative to the density matrix, is the Wigner function which is related (for continuous systems) to the density matrix representation by
\begin{equation}
W(x,p) =\frac{1}{\pi\hbar} \int_{-\infty}^{+\infty}
\, dy\, e^{2ipy/\hbar} \, \rho(x-y,x+y)
\end{equation}

It is often a useful tool to study correlations between position and momentum. If the system follows a Schr\"{o}dinger time evolution with a potential $V(x)$, the time-dependence of the Wigner function is given by the so-called \emph{Moyal bracket} $\{H,W\}_{MB}$
\begin{equation}
\frac{\partial}{\partial t} W(x,p,t) = \{H,W\}_{MB} = \{H,W\}_{PB}
+\sum_{n\geq1} \frac{\hbar^2n (-1)^n}{2^{2n}(2n+1)!} \partial_x^{2n+1} W(x,p)
\end{equation}
Where $\{H,W\}_{PB}$ is the usual Poisson braket. As in the density matrix
representation the general consequence of non unitary dynamics
(due to the environmental interactions) is the damping of oscillations
 in the Wigner function. In this representation de-coherence acts like
 diffusion in the momentum variable. In the general case the equation
 of motion of the Wigner function takes the form of a Fokker-Planck
 equation
\begin{equation}
\frac{\partial}{\partial t} W(x,p,t) = \left( -\frac{p}{m}
\frac{\partial}{\partial x} +m\omega^2 x \frac{\partial}{\partial p}
+\Lambda \frac{\partial^2}{\partial p^2} \right)\, W(x,p,t) \, .
\end{equation}

The last term is easily recognized as a diffusion term that will
generate, as in the diffusion equation, the damping of oscillations
in momentum space~\cite{PhysRevD.47.488}.

These topics are going to be further developed but the present paper do not assist theorists to complete the quantum decoherence calculus because the reference relativity cannot be developed within a unified mathematical theory due to (9).



\section{Conclusions}
The measurement problem collects the set of key questions that every
interpretation of quantum mechanics must address. This contribution
offers a new approach to the measurement problem using the concepts
shared in the information domain. The present theoretical study
yields the following results:
\begin{itemize}
\item[{-}]   The dual nature of measure;
\item[{-}] The necessary intervention of the observer;
\item[{-}] The relativistic nature of measurement.
\end{itemize}
 The
last point appears significant since elaborated procedures show how
the measurement process is the most accurate system for the
acquisition of information. Units of measurement established by
international treaties ensure a fair base to scientific knowledge.
The present theory concords with these issues through (3), (4) and
(5). Constraints 1 and 2 ensure that the measurement processes are
the most reliable and the most objective processes with respect to
other processes that acquire information, but this high position
does not imply that measurement is a zero-defects process. The
arbitrary elements (7) provoke difficulties in measuring. Physicists
are aware of these obstacles but they are reluctant to acknowledge
the systematic defects of the measurement systems due to the eminent
role of tests in the scientific method. Thinkers fear that criticism
on measurement shakes up the foundations of the scientific knowledge
and the measure relativism could deny the validity of the scientific
method. Instead this fear has not ground because (7) identifies the
arbitrary entities of a measure. The present conclusions match with
important phenomena devised by Einstein and by quantum theorists.
Points 1-5, A-C and I-IV pertain to a unified scenario. We finally
guess that the reference relativity can remove heavy prejudices upon
the mythical conception of measurement.


\begin{theacknowledgments}
  O.P. would like to acknowledge financial support from the "\textsc{Fondazione Cassa di Risparmio di Spoleto}".
\end{theacknowledgments}



\bibliographystyle{aipproc}   

\bibliography{template-8s}

\IfFileExists{\jobname.bbl}{}
 {\typeout{}
  \typeout{******************************************}
  \typeout{** Please run "bibtex \jobname" to obtain}
  \typeout{** the bibliography and then re-run LaTeX}
  \typeout{** twice to fix the references!}
  \typeout{******************************************}
  \typeout{}
 }

\end{document}